\documentclass{article}

\usepackage{arxiv}
\usepackage[utf8]{inputenc}
\usepackage[T1]{fontenc}
\usepackage[hidelinks]{hyperref}
\usepackage{url}
\usepackage{booktabs}
\usepackage{amsfonts}
\usepackage{nicefrac}
\usepackage{microtype}
\usepackage{lipsum}
\usepackage[table]{xcolor}
\usepackage{graphicx}
\usepackage{subcaption}
\usepackage{multirow}
\usepackage{makecell}
\usepackage{acronym}
\usepackage{multicol}
\usepackage{tabularx}
\usepackage{amsmath}
\usepackage{natbib}
\usepackage{float}

\graphicspath{{./images/}}

\title{Analysis of Biomarkers from Visual Explorations in Individuals Suspected to Be in the Prodromal Phase of Parkinson’s Disease}

\author{
  Maria F. Alcala-Durand \\
  Escuela T\'ecnica Superior de Ingenieros de Telecomunicaci\'on\\
  Universidad Polit\'ecnica de Madrid\\
  \texttt{maria.adurand@upm.es} \\
  \And
  J. Camilo Puerta-Acevedo \\
  Escuela T\'ecnica Superior de Ingenieros de Telecomunicaci\'on\\
  Universidad Polit\'ecnica de Madrid\\
  \texttt{juancamilo.puerta@upm.es}  \\
  \And
  Valle Pérez del Olmo \\
  Movement Disorder Unit, Neurology Service\\
  Hospital Gregorio Marañón de Madrid\\  \texttt{valle.perez@salud.madrid.org} \\
  \And
  Elisa Luque-Buzo \\
  Movement Disorder Unit, Neurology Service\\
  Hospital Gregorio Marañón de Madrid\\  \texttt{elisa.luque@salud.madrid.org} \\
  \And
  Juli\'an D. Arias-Londo\~no \\
  Escuela T\'ecnica Superior de Ingenieros de Telecomunicaci\'on\\
  Universidad Polit\'ecnica de Madrid\\
  \texttt{julian.arias@upm.es}  \\
  \And
  Francisco J. Grandas-Pérez \\
  Movement Disorder Unit, Neurology Service\\
  Hospital Gregorio Marañón de Madrid\\
  \texttt{francisco.grandas@salud.madrid.org} \\
  \And
  Juan I. Godino-Llorente \\
  Escuela T\'ecnica Superior de Ingenieros de Telecomunicaci\'on\\
  Universidad Polit\'ecnica de Madrid\\
  \texttt{ignacio.godino@upm.es} \\
}

\begin{document}
\maketitle

\begin{abstract}
Identifying prodromal Parkinson's Disease is a precondition for any future treatment that modifies the course of the disease, since such therapies would ideally act on the initial pathogenic events, early enough to delay the clinical onset of the disease rather than only slow its progression. In this work, a visual exploration model is retrained on an expanded corpus. Without any further training or threshold recalibration, it is then tested on an enriched cohort of patients with high risk of being in the prodromal phase of Parkinson's Disease, testing whether the learned signal follows a continuum of disease rather than only separating patients from controls. The resulting score distribution is analysed as an external evaluation of the hypothesis that the oculomotor signature targeted by the model is already partly present before clinical diagnosis. With every subject scored on data not used to train the model, the scores are ordered as controls $<$ high-risk patients $<$ Parkinson's Disease, with 42\% of the high-risk patients exceeding the decision threshold. A SHAP (SHapley Additive exPlanations) decomposition of the scores into classifiers and gaze features shows that high scores in this cohort are driven mainly by reduced saccadic excursion, fewer fixations and longer fixations, the hypokinetic oculomotor profile that characterises Parkinson's Disease. The model score is also compared against the prodromal markers used for recruitment. The score is not correlated with hyposmia, constipation, depression or REM sleep behaviour disorder; these results, interpreted in the context of the selection induced by the recruitment criteria, are discussed as evidence that within the high-risk cohort the score is largely not explained by the available ancillary clinical variables. Thus, the results suggest that the visual biomarkers analysed are a complementary indicator to identify patients at risk of being in the prodromal stage. 
\end{abstract}


\section*{Abbreviations}
{\small
\begin{multicols}{2}
\begin{acronym}[MDS-UPDRS]
\acro{AFT}{Average Fixation Time}
\acro{AUC}{Area Under the Receiver Operating Characteristic Curve}
\acro{BDI}{Beck Depression Inventory}
\acro{ECG}{Electrocardiogram}
\acro{EoS}{Entropy of States}
\acro{FFS}{Forward Feature Selection}
\acro{FO}{Fractional Occupancy}
\acro{GMM}{Gaussian Mixture Model}
\acro{HC}{Healthy Control}
\acro{HDA}{High-Density Area}
\acro{HGUGM}{Hospital General Universitario Gregorio Mara\~n\'on}
\acro{HUF}{Hospital Universitario de Fuenlabrada}
\acro{LD}{Longest Diagonal}
\acro{MDS-UPDRS}{Movement Disorder Society - Unified Parkinson's Disease Rating Scale}
\acro{MIL}{Mean Interval Length}
\acro{MLT}{Mean Lifetime}
\acro{MoCA}{Montreal Cognitive Assessment}
\acro{MoE}{Mixture of Experts}
\acro{PD}{Parkinson's Disease}
\acro{PHR}{Patients at High Risk of being in the Prodromal Stage of Parkinson's Disease}
\acro{RBD}{REM sleep Behaviour Disorder}
\acro{RBF}{Radial Basis Function}
\acro{SHAP}{SHapley Additive exPlanations}
\acro{SVM}{Support Vector Machine}
\acro{TF}{Total Fixations}
\acro{TS}{Total Saccades}
\acro{TSA}{Total Scanned Area}
\acro{TSE}{Total Saccadic Excursion}
\acro{UPSIT}{University of Pennsylvania Smell Identification Test}
\acro{yHC}{Young Healthy Control}
\end{acronym}
\end{multicols}
}

% =================================================================
\section{Introduction}
% =================================================================

\ac{PD} is the second most prevalent neurodegenerative disorder~\cite{liao2024deciphering} and the fastest growing neurological disorder worldwide~\cite{dorsey2018gbd}, and its diagnosis still relies on cardinal motor signs such as bradykinesia, rigidity and rest tremor~\cite{postuma2015mdscriteria}. By the time these are present, a substantial fraction of nigrostriatal dopaminergic neurons has been lost and the disease has been progressing for years or even decades, during a prodromal phase in which non-motor symptoms, subtle motor changes and biomarker abnormalities can already be detected but are insufficient for diagnosis~\cite{ugrumov2020earlydiagnosis,mahlknecht2015concept,postuma2012howdoesparkinsonismstart}. The Movement Disorder Society research criteria operationalise this stage as a probabilistic combination of risk markers, including \ac{RBD}, hyposmia, autonomic dysfunction, depression and family history of \ac{PD}~\cite{berg2015mdsresearch,heinzel2019update}; isolated \ac{RBD}, for instance, converts to \ac{PD}, dementia with Lewy bodies or multiple system atrophy at approximately 6\% per year~\cite{postuma2019rbdrisk}.

Identifying prodromal \ac{PD} is a precondition for any future treatment that modifies the course of the disease, since such therapies would ideally act early enough to delay the clinical onset of the disease~\cite{mahlknecht2015concept,ugrumov2020earlydiagnosis}, but doing so reliably and at scale remains an open problem. Individual markers are moderately predictive at best: hyposmia or constipation are sensitive but have low specificity, \ac{RBD} is specific but insensitive, and confirmatory tools such as polysomnography or DaTscan are expensive, invasive, or both~\cite{heller2017rbdimaging,mitchell2021emerging,heinzel2019update}. Noninvasive, low burden modalities that complement existing markers are therefore needed; accelerometry recorded with wearable devices, for instance, can identify individuals who will be diagnosed with \ac{PD} up to seven years in advance~\cite{schalkamp2023wearable}.

Eye movements are a promising candidate: oculomotor control depends on the basal ganglia, brainstem and frontal cortical circuits affected by \ac{PD}, and eye movement abnormalities in \ac{PD} are well documented in both controlled and naturalistic tasks~\cite{antoniades2024eye,kassavetis2022eye,rodriguez2019eye,reiner2023oculometric,ko2021abnormal,culicetto2025eyetracking}. In free visual exploration, \ac{PD} patients show fewer and shorter saccades, longer fixations and a reduced spatial extent of gaze~\cite{matsumoto2011small,wong2018eye,wong2020prolonged,dietz2011emotion}. These abnormalities are not restricted to clinically diagnosed \ac{PD}: subtle changes in saccades, pupils and blinking have been reported in idiopathic \ac{RBD}~\cite{hanuska2019eyemovements,perkins2021rbdoculomotor,habibi2022eyetracking,habibi2025saccade}, in non-manifesting carriers of LRRK2 mutations~\cite{riek2024lrrk2oculomotor}, in prodromal X-linked dystonia-parkinsonism~\cite{mertin2023xdp} and in unaffected first-degree relatives of \ac{PD} patients~\cite{blekher2009multiplesteppattern}, and oculomotor measures appear to follow the disease continuum from healthy controls through prodromal stages to clinical \ac{PD}~\cite{brien2023classification,turcano2019early,lachwojnarowicz2025ophthalmicbiomarkers}.

Machine learning models that produce a continuous score of \ac{PD} likelihood are particularly well suited to this question: rather than asking whether a subject crosses a diagnostic boundary, one can ask whether the score lies between the healthy and clinical distributions. Such sensitivity to the disease stage has been shown for a saccadic task across stages of \ac{PD}~\cite{brien2023classification}, and for clinical scales~\cite{skorvanek2017differences} and \ac{ECG}-based biomarkers~\cite{karabayir2023ecg}.

In a previous work~\cite{alcala2025screening}, we introduced an interpretable framework for the automatic screening of \ac{PD} from free visual explorations of six images, a task that requires only minimal instructions to the participant. It combines classical fixation and saccadic metrics with features of \acp{HDA}, defined as \ac{GMM} components fitted to the densest areas of the explored regions; one classifier per image, applied to each eye separately, produces a probability per eye, and a \ac{MoE} integrates these probabilities into a score per patient. On a test set, it discriminated well between \ac{PD} patients and age-matched \acp{HC}. Whether this behaviour extends to subjects who have not yet crossed the diagnostic threshold remained an open question; the present work therefore challenges the previously derived model with an enriched cohort at risk of being in the prodromal phase, testing whether the learned signal follows a continuum of disease rather than only separating patients from controls.

The corpus has been expanded since~\cite{alcala2025screening} with more \acp{HC} and \ac{PD} patients and, more importantly, with 88 \ac{PHR}, recruited at the same hospitals with the same recording protocol and following established criteria for prodromal \ac{PD} (a first-degree relative with \ac{PD} and either hyposmia, \ac{RBD}, or at least three minor risk factors~\cite{berg2015mdsresearch,heinzel2019update}). Since, in the absence of biomarkers of the underlying synucleinopathy~\cite{hoglinger2024synneurge,simuni2024nsdiss}, membership of the prodromal stage can only be established after phenoconversion~\cite{berg2014redefine}, these participants are referred to as being at high risk of being in the prodromal stage rather than as prodromal \ac{PD} cases.

The main contribution of this work is the external evaluation of a \ac{PD} visual exploration model on a cohort of \ac{PHR} subjects, a population that remains underrepresented in the eye tracking literature. The retrained model is applied without any further training or threshold recalibration to the \ac{PHR} cohort, held out from every stage of model development, as a test of whether the oculomotor signature targeted by the model is already partly present before clinical diagnosis. Supporting this, a \ac{SHAP} analysis decomposes each score into contributions of the classifiers of each exploration and eye and of the gaze features within them, enabling explanations both for the population and for each subject. The model score is also compared against the prodromal markers used for recruitment, to assess whether it carries complementary information. To enable all of the above, the pipeline from~\cite{alcala2025screening} is retrained on an enlarged training set with its hyperparameters reselected by the same cross-validation protocol and its architecture unchanged.

% =================================================================
\section{Materials}
\label{sec:materials}
% =================================================================
This section presents the corpus of participants with a special emphasis on the different cohorts recorded, the recruitment criteria followed, and the recording setup used throughout the paper. 

\paragraph{Participants.}
The corpus comprises four main cohorts: clinically diagnosed \ac{PD} patients, young \acp{HC}, age-matched \acp{HC}, and \ac{PHR}, all recruited at \ac{HUF} and \ac{HGUGM} in Madrid, Spain. Of the 273 participants included in the study database (85 \ac{PD}, 79 \ac{HC}, 21 \ac{yHC} and 88 \ac{PHR}), 260 had recordings that passed the quality control described below and were analysed: 78 \ac{PD}, 73 \ac{HC}, 21 \ac{yHC} and 88 \ac{PHR}. \ac{PD} patients had less than 5 years of disease evolution and were assessed in the ON medication state. \ac{MDS-UPDRS} Part III and \ac{MoCA} scores were collected for the \ac{PD} cohort, and \ac{MoCA} was additionally administered to \ac{HC} and \ac{yHC}.

\paragraph{\ac{PHR} cohort and recruitment criteria.}
\ac{PHR} subjects were recruited following a protocol designed to identify an enriched population, consistent with the Movement Disorder Society research criteria for prodromal \ac{PD}~\cite{berg2015mdsresearch,heinzel2019update}. Eligibility required a first-degree relative with \ac{PD} (asymptomatic carriers of pathogenic monogenic \ac{PD} variants were also eligible), age over 50 years, and either at least one major marker, namely hyposmia or anosmia confirmed by the \ac{UPSIT}~\cite{doty1984upsit} or \ac{RBD} clinically suggested or confirmed by polysomnography, or at least three of five minor criteria: not smoking; low caffeine consumption (fewer than three cups of coffee or six cups of tea per week); depressive symptoms assessed by the \ac{BDI}~\cite{beck1961inventory}; constipation requiring treatment more than once per week or fewer than one spontaneous bowel movement per day; and male sex. All clinical assessments were conducted by the specialist physicians of the recruiting hospitals, and all \ac{PHR} participants underwent the same eye tracking protocol as the other cohorts. By construction, the \ac{PHR} cohort is enriched in subjects carrying multiple prodromal markers, which affects the interpretation of the complementarity analysis (Section~\ref{sec:complementarity}).

\paragraph{Recording and preprocessing.}
Stimuli, recordings and preprocessing follow the protocol of~\cite{alcala2025screening} without modification. Eye movements were captured with a binocular EyeLink\textsuperscript{®}~1000~Plus eyetracker at 1~kHz~\cite{srresearch2017eyelink} while participants explored six images (a circle, a cube, a house, a pair of intersecting pentagons, the Rey-Osterrieth complex figure~\cite{osterrieth1944test}, and a clock with numbers and hands) for 15~s each, in a fixed order and under homogeneous environmental conditions. Samples tagged as blinks were excluded, fixations and saccades were extracted with the eyetracker's online parsing system, gaze coordinates were normalised preserving the aspect ratio of the screen, and observations with insufficient data, excessive blinking or inter-eye dissimilarity were excluded.

\paragraph{Data partition.}
For the binary \ac{PD} vs. control task, in which \ac{HC} and \ac{yHC} participants form the control class (in~\cite{alcala2025screening}, \ac{yHC} participants were used only descriptively), 20\% of the participants of each class were randomly sampled at the subject level, with a fixed seed and before any feature extraction, as an internal test set, excluded from all training and selection stages. After quality control, the training set comprised 138 participants (63 \ac{PD}, 59 \ac{HC} and 16 \ac{yHC}) and the test set 34 (15 \ac{PD}, 14 \ac{HC} and 5 \ac{yHC}). The \ac{PHR} cohort was not used at any stage of model development.

% =================================================================
\section{Methods}
\label{sec:methods}
% =================================================================

The architecture (feature extraction, one classifier per exploration and \ac{MoE} aggregator) is that of~\cite{alcala2025screening}, to which the reader is referred for the mathematical details; this section focuses on what is new or was re-evaluated with the expanded corpus.

\subsection{Feature Extraction}

For each visual exploration and each eye, the two feature sets defined in \cite{alcala2025screening} were extracted: (i) the six classical fixation and saccade metrics (\ac{TS}, \ac{TSE}, \ac{TF}, \ac{AFT}, \ac{TSA}, and \ac{LD}); and, (ii) the \ac{HDA}-based features obtained by fitting a \ac{GMM} with $k$ components to the gaze data of the control participants of the training set and computing, for each component, \ac{FO}, \ac{MLT}, \ac{MIL}, and \ac{EoS}, complemented by the global maxima and minima of these descriptors across components. Concatenating both sets yields the per-exploration, per-eye feature vector. The number of components $k$ is treated as a hyperparameter selected via cross-validation, as in \cite{alcala2025screening}.

\subsection{Model training and selection}

\begin{figure}[t!]
\centering
\includegraphics[width=\linewidth]{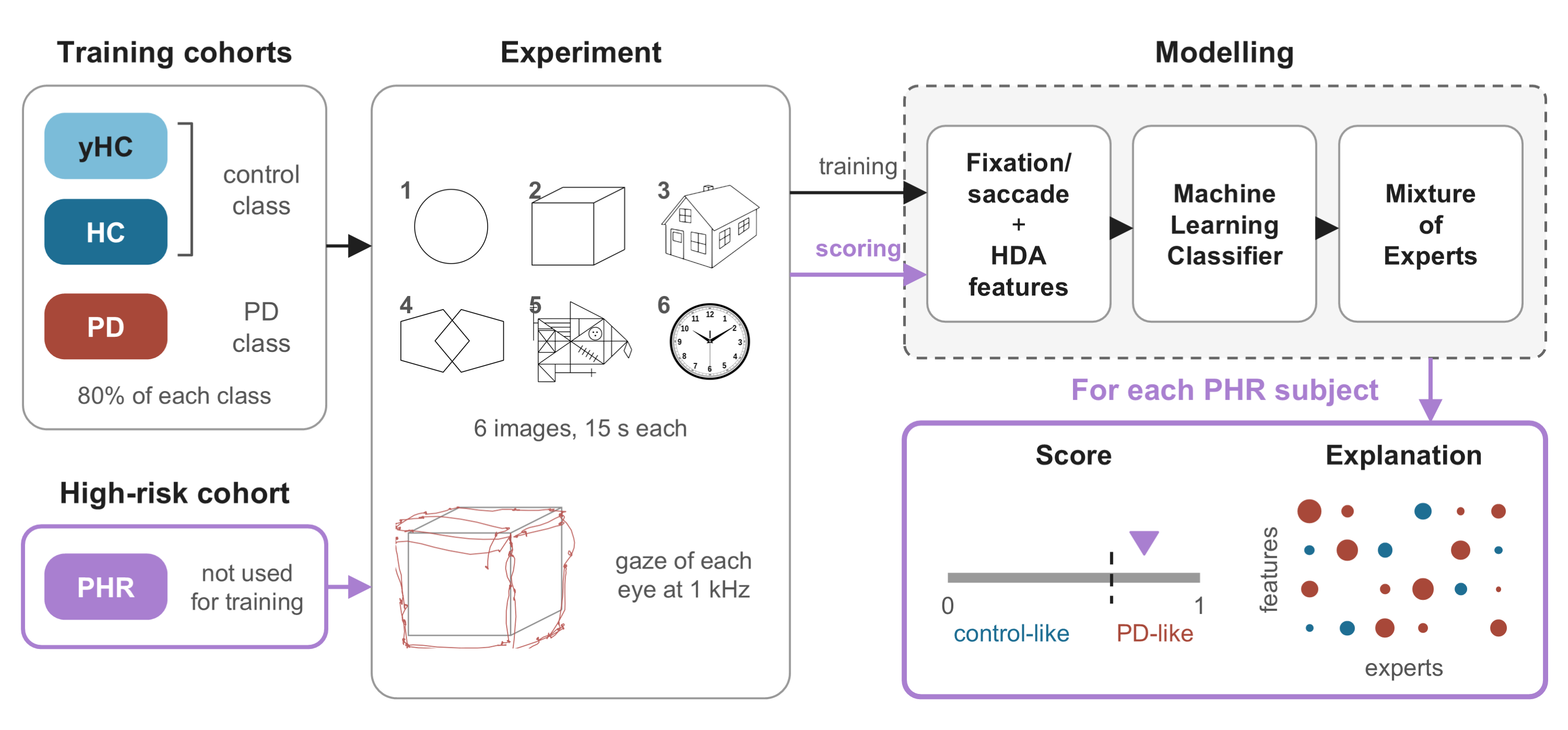}
\caption{Overview of the study. The model is trained on the \ac{yHC}, \ac{HC} (control class) and \ac{PD} participants of the training set: fixation/saccade and \ac{HDA}-based features are extracted for each exploration and eye, one \ac{SVM} per exploration scores each eye, and the \ac{MoE} combines the selected experts into a patient-level score. The \ac{PHR} participants, not used in model development, are scored by the trained model, and each score is explained in terms of experts and features.}
\label{fig:pipeline}
\end{figure}

For each of the six explorations, a single classifier was trained on the feature vectors of both eyes and applied to each eye separately, yielding a calibrated probability of \ac{PD} for each of the twelve exploration-eye combinations, hereafter referred to as experts~\cite{bishop2006pattern,hastie2009elements}. A second \ac{MoE} stage integrates the probabilities of the experts retained by \ac{FFS} into a score per patient~\cite{kittler1998combining,sharkey1996combining,zhou2012ensemble,wolpert1992stacked}, which is compared with a decision threshold (Fig.~\ref{fig:pipeline}). The pipeline was retrained from scratch on the expanded training set with the cross-validation protocol of~\cite{alcala2025screening} (20 random splits at the subject level), maximising Youden's J. All hyperparameters were reselected, namely the parameters $C$ and $\gamma$ of the \ac{SVM} with \ac{RBF} kernel (shared by the six explorations), the number of \ac{GMM} components $k$ per exploration, the \ac{MoE} aggregation rule, the retained experts and the threshold. The model families are unchanged. At these sample sizes, the approximately twofold increase in the size of the training set is a substantial gain of information, so the hyperparameters were refitted rather than carried over from~\cite{alcala2025screening}, where they were estimated on a much smaller training sample.

\subsection{External evaluation and complementarity analyses}

The trained pipeline (\ac{GMM} components, \acp{SVM}, \ac{MoE} aggregation rule and threshold) was applied to each \ac{PHR} subject without any further training, refitting or threshold recalibration. All cohort comparisons use scores obtained on data not used to fit the model. The training subjects were scored by repeated 10-fold cross-validation, and the test and \ac{PHR} subjects by the trained model. The ordering controls (\ac{HC} and \ac{yHC})~$<$~\ac{PHR}~$<$~\ac{PD} was tested with the Jonckheere-Terpstra test, complemented by pairwise Mann-Whitney $U$ tests. As effect size, the probability that a randomly chosen subject of one cohort scores higher than a randomly chosen subject of the other is reported (0.5: no difference). Within the \ac{PHR} cohort, Spearman correlations were computed between the score and the markers used for recruitment (hyposmia, \ac{RBD}, depression and constipation) and among the markers, and the scores of the subjects with corroborating findings of nigrostriatal dysfunction from follow-up (altered substantia nigra echography or pathological DaTscan) were compared with those of the remaining subjects with Mann-Whitney $U$ tests.

\subsection{SHAP interpretability at two levels}

\ac{SHAP} attributes a prediction to the input features by distributing the difference between the prediction and a baseline according to Shapley values~\cite{lundberg2017unified}. At the expert level, Kernel \ac{SHAP} was applied to each retained expert, with a background of 80 training subjects, giving the contribution $\phi_{e,f}(\mathbf{x})$ of each fixation/saccade or \ac{HDA}-based feature $f$ to the probability $p_e(\mathbf{x})$ of expert $e$. At the \ac{MoE} level, the aggregator is a weighted average of the probabilities of the retained experts, $p(\mathbf{x}) = \sum_e w_e\, p_e(\mathbf{x})$, with weights proportional to the cross-validated \ac{AUC} of their exploration and summing to one; since, by the efficiency property of \ac{SHAP}, $p_e(\mathbf{x}) - \mathbb{E}[p_e] = \sum_f \phi_{e,f}(\mathbf{x})$, the two levels combine exactly:
\begin{equation}
p(\mathbf{x}) = \sum_e w_e\, \mathbb{E}[p_e] + \sum_e \sum_f w_e\, \phi_{e,f}(\mathbf{x}),
\label{eq:decomposition}
\end{equation}
that is, each feature of each expert contributes $c_{e,f}(\mathbf{x}) = w_e\, \phi_{e,f}(\mathbf{x})$ to the patient-level score. The analysis was restricted to the out-of-sample subjects (internal test set and \ac{PHR} cohort). For each feature and expert, importance is the mean of $|c_{e,f}|$ and direction is the Spearman correlation between the feature value and $c_{e,f}$ (positive: higher values push the score towards \ac{PD}). The net contribution of a feature, $C_f(\mathbf{x}) = \sum_e c_{e,f}(\mathbf{x})$, sums over all experts for the fixation/saccade features, which have the same meaning in every exploration, and over the two eyes of an exploration for the exploration-specific \ac{HDA}-based features; global importance is the mean of $|C_f|$. Mean contributions were also computed for the controls and \ac{PD} patients of the test set and for the \ac{PHR} subjects below and above the threshold. All analyses used Python with \texttt{scikit-learn} (v1.8.0).

% =================================================================
\section{Results}
\label{sec:results}
% =================================================================

This section first describes the fixation/saccade features across the four cohorts, and then reports the performance of the retrained pipeline on the \ac{PD} vs. control task, its application to the \ac{PHR} cohort, the interpretability analysis, and the relation between the model score and the prodromal markers used for recruitment.

\subsection{Progression of the features across cohorts}

\begin{figure}[h]
\centering
\includegraphics[width=\linewidth]{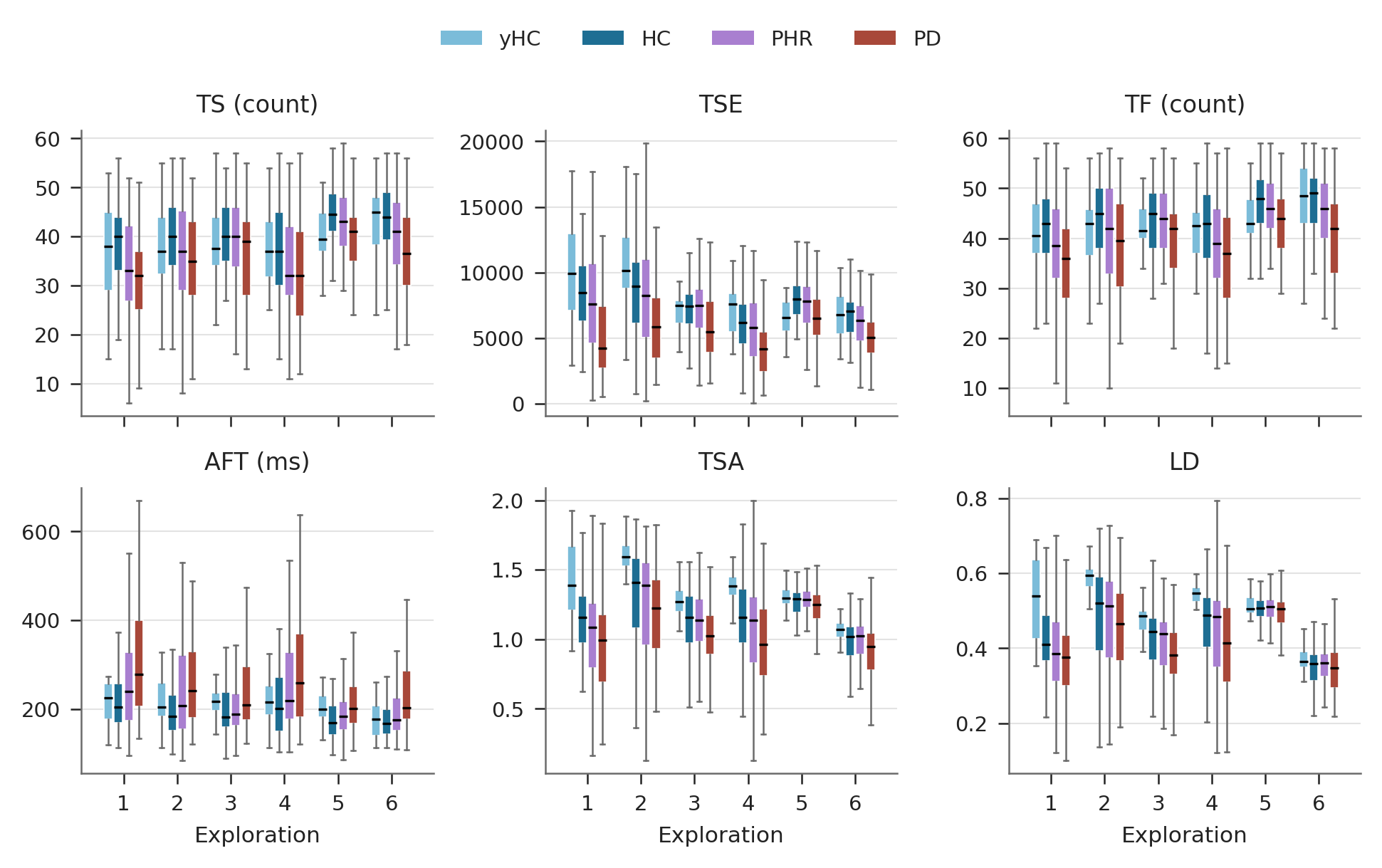}
\caption{Distribution of fixation/saccade-based metrics across the four cohorts (\ac{yHC}, \ac{HC}, \ac{PHR}, \ac{PD}) for each of the six visual explorations. For most metrics and most explorations, the \ac{PHR} box sits between \ac{HC} and \ac{PD}, with \ac{PD} displaying the most affected values.}
\label{fig:boxplots_sacfix}
\end{figure}

Fig.~\ref{fig:boxplots_sacfix} shows the six fixation/saccade-based features of the four cohorts in each exploration.

In all six explorations, \ac{PD} subjects showed the lowest \ac{TS}, \ac{TSE}, \ac{TF}, \ac{TSA} and \ac{LD} and the highest \ac{AFT}, consistent with the hypokinetic profile reported in~\cite{matsumoto2011small,wong2018eye,wong2020prolonged,alcala2025screening}, and the \ac{PHR} median lay between the \ac{HC} and \ac{PD} medians in 32 of the 36 combinations of metric and exploration: the intermediate position of the \ac{PHR} cohort is already visible in the raw features, before any model is applied.

\subsection{Retrained pipeline performance}

The selected configuration and the cross-validated \ac{AUC} of each exploration are reported in Table~\ref{tab:retrained_hparams}. Individually, the most discriminative exploration was the Rey-Osterrieth complex figure (Expl.~5; \ac{AUC} $0.75$), followed by the circle (Expl.~1; $0.73$), and the least discriminative was the house (Expl.~3; $0.68$). The \ac{FFS} retained four experts spanning three images, and the selected aggregator was, as in~\cite{alcala2025screening}, an \ac{AUC}-weighted voting scheme, i.e., the average of their probabilities weighted by the cross-validated \ac{AUC} of their exploration, with a decision threshold of 0.45.

\begin{table}[h]
\centering
\caption{Selected configuration per exploration: number of \ac{GMM} components $k$ ($C = 1$ and $\gamma = 0.001$ for all explorations), cross-validated \ac{AUC} (mean $\pm$ std) and experts retained by the \ac{FFS} (L/R: left/right eye).}
\label{tab:retrained_hparams}
\begin{tabular}{lccc}
\toprule
Exploration & $k$ & CV \ac{AUC} & Experts \\
\midrule
Expl.~1 & 12 & $0.73 \pm 0.09$ & R \\
Expl.~2 & 17 & $0.70 \pm 0.07$ & L \\
Expl.~3 & 16 & $0.68 \pm 0.09$ & -- \\
Expl.~4 & 49 & $0.69 \pm 0.06$ & -- \\
Expl.~5 & 13 & $0.75 \pm 0.07$ & L, R \\
Expl.~6 & 24 & $0.70 \pm 0.09$ & -- \\
\bottomrule
\end{tabular}
\end{table}

Under cross-validation, the \ac{MoE} achieved an \ac{AUC} of $0.77 \pm 0.07$ (sensitivity $0.77 \pm 0.12$, specificity $0.73 \pm 0.12$, F1 $0.73 \pm 0.09$). On the test set (15 \ac{PD}, 14 \ac{HC} and 5 \ac{yHC}), it achieved an \ac{AUC} of $0.76$ (sensitivity $0.87$, specificity $0.53$, F1 $0.70$). On the training subjects, the \ac{AUC} of the cross-validated scores used below was $0.79$.

\subsection{External evaluation on the \texorpdfstring{\ac{PHR}}{PHR} cohort}

\begin{figure}[h]
\centering
\includegraphics[width=0.55\linewidth]{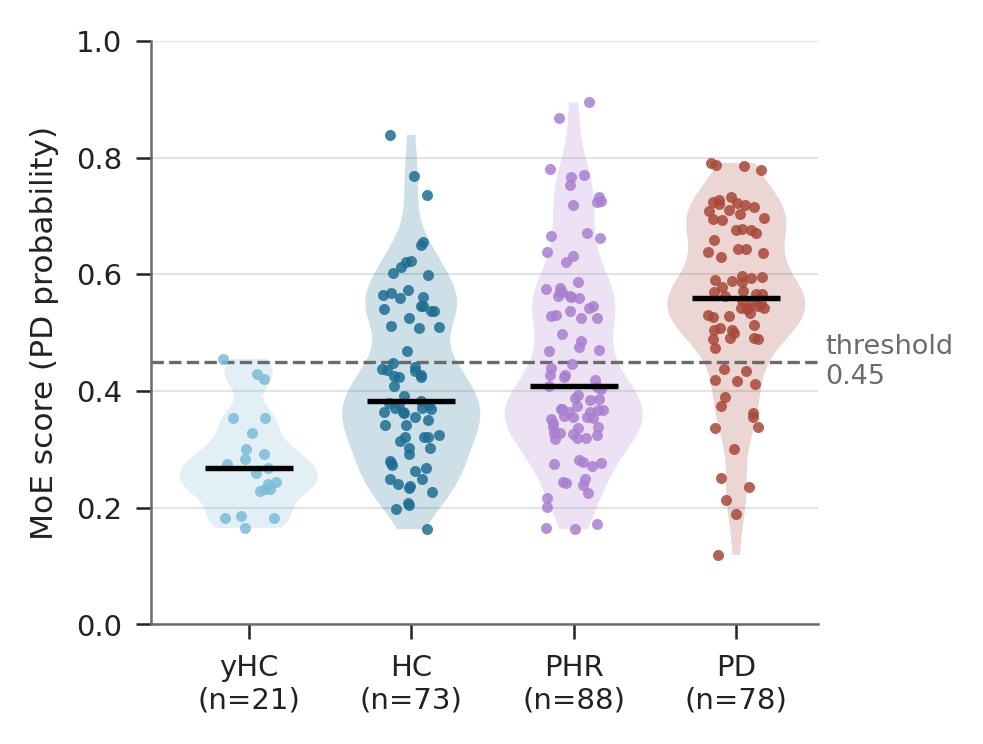}
\caption{\ac{MoE} scores of the four cohorts, obtained on data not used to fit the model (cross-validated for the training subjects). Black bars: medians; dashed line: decision threshold.}
\label{fig:prob_raincloud_post_moe}
\end{figure}

Fig.~\ref{fig:prob_raincloud_post_moe} shows the \ac{MoE} scores of the four cohorts. The median score increases from the controls ($0.36$; $0.27$ for \ac{yHC} and $0.38$ for \ac{HC}) to \ac{PHR} ($0.41$) and \ac{PD} ($0.56$), a stochastic ordering confirmed by the Jonckheere-Terpstra test (\textit{JT}~$= 15460$, \textit{z}~$= 6.41$, \textit{p}~$< 10^{-10}$). \ac{PHR} subjects scored higher than the controls and lower than \ac{PD}. A randomly chosen \ac{PHR} subject scored higher than a randomly chosen control with probability $0.61$ (\textit{p}~$= 0.01$) and lower than a randomly chosen \ac{PD} patient with probability $0.69$ (\textit{p}~$< 10^{-4}$), and a \ac{PD} patient scored higher than a control with probability $0.78$ (\textit{p}~$< 10^{-9}$). The \ac{PHR} scores overlap with those of both cohorts (interquartile range $0.33$--$0.56$, vs. $0.27$--$0.50$ for the controls and $0.49$--$0.67$ for \ac{PD}). Overall, 42\% of the \ac{PHR} subjects (37 of 88) exceeded the decision threshold of 0.45, compared with 28\% of the controls (26 of 94) and 78\% of the \ac{PD} patients (61 of 78).

\subsection{Interpretability}
\label{sec:interpretability}

The four experts have similar weights in the \ac{MoE} (0.24 to 0.26), and their mean absolute contributions to the score were also similar, from $0.043$ for Expl.~2 (left eye) to $0.048$ for Expl.~5 (left eye).

\begin{figure}[t!]
\centering
\includegraphics[width=\linewidth]{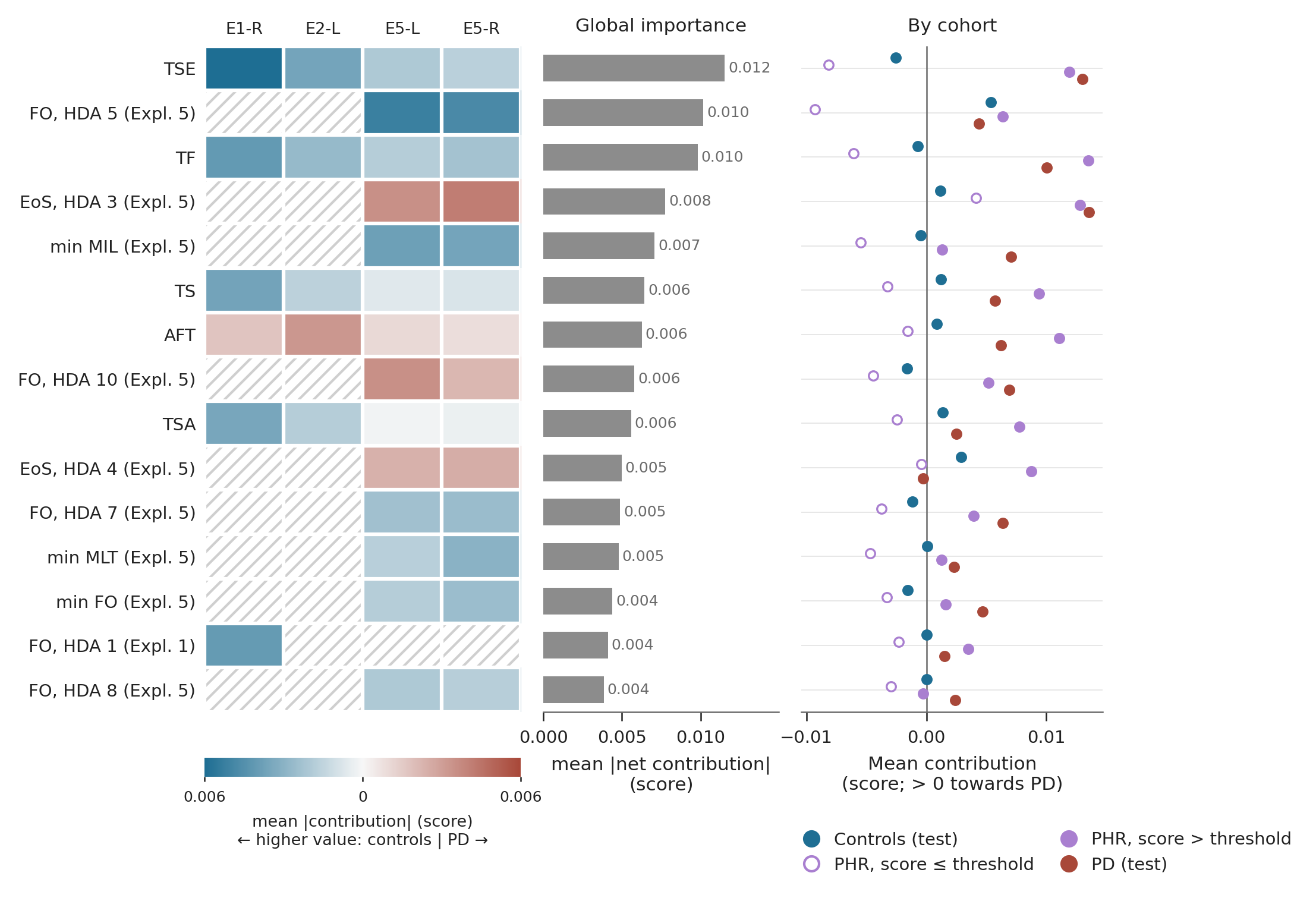}
\caption{Interpretability at the feature level for the 15 features with the largest global importance, over the subjects not used for training. Left: experts (E$n$-L/R: Expl.~$n$, left/right eye); colour intensity: mean absolute contribution of the feature to the score through that expert (Eq.~\eqref{eq:decomposition}); hue: direction (red: higher values push the score towards \ac{PD}; blue: towards the controls); hatched: feature not defined for that expert. Centre: global importance (mean absolute net contribution). Right: mean contribution for the controls and \ac{PD} patients of the test set and for the \ac{PHR} subjects below and above the threshold.}
\label{fig:feature_matrix}
\end{figure}

\begin{figure}[t!]
\centering
\includegraphics[width=0.85\linewidth]{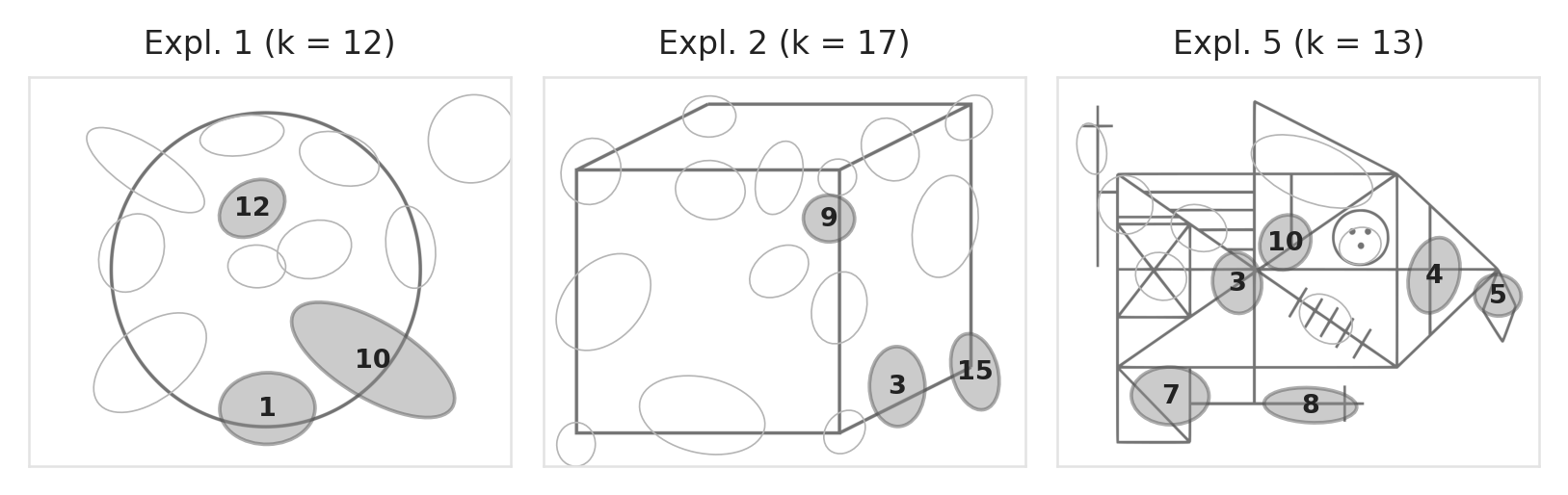}
\caption{Location of the most influential \acp{HDA} (grey, labelled) in the three explorations retained by the \ac{MoE}, namely those among the features of Fig.~\ref{fig:feature_matrix} and the three most influential \acp{HDA} of each exploration. All \ac{GMM} components are shown as ellipses of one standard deviation.}
\label{fig:hda_locations}
\end{figure}

Fig.~\ref{fig:feature_matrix} shows the 15 features with the largest global importance. \ac{TSE} is the most influential, followed by the \ac{FO} of \ac{HDA}~5 of Expl.~5, by \ac{TF} and by other \ac{HDA}-based features of Expl.~5. The directions of these fixation/saccade features are consistent across experts and agree with the hypokinetic profile of \ac{PD} (Fig.~\ref{fig:boxplots_sacfix}): lower \ac{TSE} and \ac{TF} and longer \ac{AFT} push the score towards \ac{PD} (Spearman correlations between feature value and contribution from $-0.94$ to $-0.66$ for \ac{TSE} and \ac{TF}, and from $0.45$ to $0.84$ for \ac{AFT}). The 186 \ac{HDA}-based features account for 80\% of the total attribution, and the six fixation/saccade features for the remaining 20\%. On average, a single fixation/saccade feature therefore contributes about eight times more than a single \ac{HDA}-based feature. The influential \acp{HDA} lie at specific parts of the stimuli (Fig.~\ref{fig:hda_locations}): for instance, a larger \ac{FO} of \ac{HDA}~5 of Expl.~5, at the right tip of the Rey-Osterrieth figure, pushes the score towards the controls, whereas a larger \ac{FO} of \ac{HDA}~10, near the centre of the figure, pushes it towards \ac{PD}.

Among the \ac{PHR} subjects, those above the threshold are pushed towards \ac{PD} mainly by \ac{TF}, \ac{TSE} and \ac{AFT}, with mean contributions ($0.014$, $0.012$ and $0.011$) similar to or larger than those of the \ac{PD} patients of the test set ($0.010$, $0.013$ and $0.006$), whereas the same features push those below the threshold towards the controls (Fig.~\ref{fig:feature_matrix}, right); the \ac{HDA}-based features of Expl.~5 contribute slightly more in the \ac{PD} patients. The errors on the test set were mostly driven by \ac{HDA}-based features: those of Expl.~5 for the nine false positives, and those of Expl.~5 and~1 for the two false negatives.

\subsection{Complementarity with prodromal markers}
\label{sec:complementarity}

\begin{figure}[t!]
\centering
\includegraphics[width=0.5\linewidth]{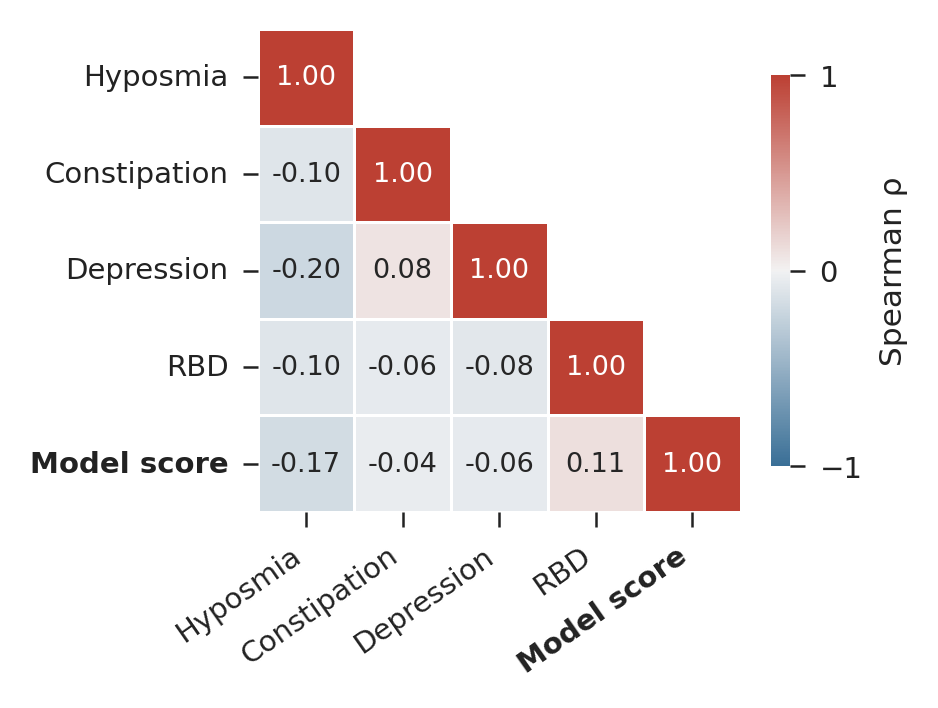}
\caption{Spearman correlations among the four markers used for recruitment and the model score within the \ac{PHR}.}
\label{fig:prodromal_corr}
\end{figure}

Among the 88 \ac{PHR} subjects, hyposmia (\ac{UPSIT}~\cite{doty1984upsit}) was present in 26\%, constipation in 34\%, depressive symptoms (\ac{BDI}~\cite{beck1961inventory}) in 28\% and \ac{RBD} in 34\% (Fig.~\ref{fig:prodromal_corr}).

The markers were weakly correlated with one another (all $|\rho| \le 0.20$, none significant), as expected from a heterogeneous prodromal phenotype~\cite{mahlknecht2015concept,heller2017rbdimaging} and from the recruitment criteria, under which only the subjects without hyposmia or \ac{RBD} needed three minor criteria: all the correlations between a major and a minor marker were negative, as was that between hyposmia and \ac{RBD}. The model score was not correlated with any of the markers ($|\rho| \le 0.17$, all $p \ge 0.12$). The scores of the 32 \ac{PHR} subjects with corroborating findings of nigrostriatal dysfunction did not differ significantly from those of the remaining 56 (medians $0.38$ and $0.42$; $p = 0.49$).

% =================================================================
\section{Discussion}
\label{sec:discussion}
% =================================================================

The main result of this work is that a visual exploration model trained to distinguish \ac{PD} from controls, applied without any further training or threshold recalibration to a \ac{PHR} cohort held out from model development, produces scores ordered as controls~$<$~\ac{PHR}~$<$~\ac{PD}. This is not a classification experiment on the \ac{PHR} cohort, but an external evaluation of the hypothesis that the oculomotor signature targeted by the model is already partly present before clinical diagnosis.

This is consistent with the intermediate oculomotor profiles reported in idiopathic \ac{RBD}~\cite{hanuska2019eyemovements,perkins2021rbdoculomotor,habibi2022eyetracking,habibi2025saccade}, in non-manifesting LRRK2 carriers~\cite{riek2024lrrk2oculomotor} and in unaffected first-degree relatives of \ac{PD} patients~\cite{blekher2009multiplesteppattern}, and extends them to free visual exploration, one of the simplest and least demanding eye tracking tasks, which requires only minimal instructions.

The \ac{PHR} scores span a wide range ($0.16$--$0.89$) that overlaps with both the controls and \ac{PD}, and more \ac{PHR} subjects than controls exceed the threshold (42\% vs. 28\%), as expected from a cohort at risk in which the score may act as a complementary indicator of risk rather than as a binary flag. The decomposition at two levels makes this heterogeneity clinically inspectable: for any subject, the attending physician can see which experts and which gaze behaviours drove the score. At the population level, \ac{PHR} subjects with high scores share the profile that the model associates with \ac{PD}, namely reduced saccadic excursion, fewer fixations and longer fixations, whose association with \ac{PD} is well established in free viewing tasks~\cite{matsumoto2011small,wong2018eye,wong2020prolonged}, whereas the \ac{HDA}-based features of the most complex stimulus contribute slightly less than in \ac{PD}; the signal detected in the \ac{PHR} cohort thus corresponds mainly to global oculomotor slowing, with a smaller contribution of exploration strategies specific to each stimulus.

The model score was unrelated to the recruitment defining markers. By construction, every \ac{PHR} subject presents hyposmia, \ac{RBD} or at least three minor criteria, which restricts and structures the variability of the markers within the cohort, and the weak correlations among them~\cite{mahlknecht2015concept,heller2017rbdimaging} indicate that no single marker is a strong proxy for the model output. Within this enriched cohort, the visual exploration score therefore captures variability that is not explained by the available markers, and is a complementary rather than redundant candidate for the screening of patients at risk of being in the prodromal stage.

\paragraph{Limitations.}
The \ac{PHR} cohort is cross-sectional, so whether high scores anticipate conversion to \ac{PD} can only be established with the longitudinal follow-up of the cohort, which is ongoing. The \ac{PHR} participants were also recruited with specific inclusion criteria, so their score distribution may differ in atrisk populations defined by other criteria.

% =================================================================
\section{Conclusions}
\label{sec:conclusions}

This study provides evidence that an interpretable machine learning model trained on free visual exploration tasks captures an oculomotor signature that reflects the continuum of neurodegenerative progression associated with \ac{PD} rather than merely acting as a binary classifier. When externally evaluated on an enriched cohort of \ac{PHR} --without any model retraining, parameter tuning, or threshold recalibration-- the generated scores exhibited a clear and statistically significant stochastic ordering ($\text{HC} < \text{PHR} < \text{PD}$). Notably, 42\% of the high-risk subjects exceeded the decision threshold established for clinical \ac{PD}, suggesting that subtle oculomotor alterations precede the onset of cardinal motor symptoms. \ac{SHAP} interpretability analyses further confirmed that high scores within the PHR cohort were predominantly driven by a generalized oculomotor slowing—specifically reduced total saccadic excursion, fewer fixations, and prolonged fixation durations --which directly mirrors the hypokinetic gaze profile characteristic of diagnosed \ac{PD} patients.   

Importantly, the visual exploration score showed no significant correlation with the ancillary clinical and risk markers used for participant recruitment, including hyposmia, constipation, depressive symptoms, and REM sleep behaviour disorder. In the context of an enriched high-risk cohort, this lack of correlation indicates that visual biomarkers capture distinct physiological variance not accounted for by traditional screening tools, offering complementary and non-redundant diagnostic information. While longitudinal follow-up remains necessary to confirm the rate of phenoconversion among high-scoring individuals, these findings validate free visual exploration as a rapid, non-invasive, and low-burden digital biomarker. Ultimately, results suggest the plausibility of integrating eye-tracking analytics into existing multi-modal risk stratification frameworks holds substantial potential to enhance the early detection of prodromal \ac{PD}, opening crucial windows for timely neuroprotective interventions.  

\section*{Acknowledgments}
This work was supported by the Ministry of Economy and Competitiveness of Spain under Grant PID2021-128469OB-I00, by Ministry of Health of Spain under Grant
FIS PI21/00771, and by Comunidad de Madrid, Spain. Universidad Politécnica de Madrid supports Julián D. Arias-Londoño through the María Zambrano grant UP2021-035, funded by the European Union--NextGenerationEU. This is also supported by a grant of the MISTI MIT Global Experiences. 
The authors would also like to thank the Madrid ELLIS unit (European Laboratory for Learning \& Intelligent Systems) for its indirect support.
Finally, the authors acknowledge Universidad Politécnica de Madrid for providing computing resources on the Magerit Supercomputer.

\subsection*{Ethics Declaration}
The study was approved by the Ethics Review Board of the \ac{HUF} and \ac{HGUGM} with codes 18/11-ENM1 and 11/2015 respectively, and in accordance with the Spanish Ethical Review Act.

\section*{Declaration of generative AI and AI-assisted technologies in the writing process}

During the preparation of this manuscript, the authors used commercially available large language models to improve the English language and enhance the clarity and style of the text. Following the use of these tools, the authors carefully reviewed and edited the content as necessary and assume full responsibility for the accuracy, integrity, and content of the published work.

\section*{Declaration of competing interest}
The authors declare that they have no competing financial interests or personal relationships that could be perceived as having influenced the research, authorship, or publication of this article.

%\newpage
%\appendix
%\section{Appendix}
%\label{app:appendix}

\bibliographystyle{unsrt}
\bibliography{references}

\end{document}